\RequirePackage[2020-02-02]{latexrelease}
\documentclass[prl,aps,twocolumn,amsmath,amssymb,superscriptaddress]{revtex4}
\usepackage{braket}
\usepackage{adjustbox}
\usepackage{float}
\usepackage{amsmath}
\usepackage{amssymb}
\usepackage{amsfonts}
\usepackage{euscript}
\usepackage{enumerate}
\usepackage{hhline}
\usepackage{pslatex}
\usepackage{tabularx}
\usepackage[usenames,dvipsnames]{xcolor}
\usepackage{threeparttable}
\usepackage{graphicx}
\usepackage{dcolumn}
\usepackage{bm}
\usepackage[sort&compress]{natbib}
\makeatletter

\renewcommand{\@biblabel}[1]{#1. }
\renewcommand{\@dotsep}{500}
\renewcommand{\@pnumwidth}{0em}
\renewcommand{\l@figure}[2]{
\@dottedtocline{1}{1.5em}{2em}{Figure #1}{}\vspace{15pt}}

\begin{document}

\title{Super quasi-bound state in the continuum}

\author{Zhanyuan Zhang}
\thanks{These authors contributed equally to this work}
\affiliation{Guangdong Provincial Key Laboratory of Information Photonics Technology, Institute of Advanced Photonic Technology, School of Information Engineering, Guangdong University of Technology, Guangzhou, China}

\author{Evgeny Bulgakov}
\thanks{These authors contributed equally to this work}
\affiliation{Kirensky Institute of Physics Federal Research Center KSC SB RAS 660036 Krasnoyarsk Russia}

\author{Konstantin Pichugin}
\affiliation{Kirensky Institute of Physics Federal Research Center KSC SB RAS 660036 Krasnoyarsk Russia}

\author{Almas Sadreev}
\email{almas@tnp.krasn.ru}
\affiliation{Kirensky Institute of Physics Federal Research Center KSC SB RAS 660036 Krasnoyarsk Russia}

\author{Yi Xu}
\email{yixu@gdut.edu.cn}
\affiliation{Guangdong Provincial Key Laboratory of Information Photonics Technology, Institute of Advanced Photonic Technology, School of Information Engineering, Guangdong University of Technology, Guangzhou, China}

\author{Yuwen Qin}
\affiliation{Guangdong Provincial Key Laboratory of Information Photonics Technology, Institute of Advanced Photonic Technology, School of Information Engineering, Guangdong University of Technology, Guangzhou, China}
\affiliation{Southern Marine Science and Engineering Guangdong Laboratory, Zhuhai 519000, China}

\date{\today}

\begin{abstract}
\noindent Avoided crossing of resonances and merging multiple bound states in the continuum (BICs) are parallel means for tailoring the physical properties of BICs. Herein, we introduce a new concept of super quasi-BIC for photonic crystal (PhC) systems where its quality ($Q$) factor is boosted in both parametric and momentum spaces. A super quasi-BIC with substantial enhancement of $Q$ factor can be achieved in a finite PhC by combining avoiding crossing of two symmetry protected (SP) quasi-BICs in parametric space and merging BICs in momentum space simultaneously. More importantly, analytical theory shows that the proposed mechanism results in the transition of asymptotic behavior of the $Q$ factor over the numbers of resonators from $N^2$ to exclusive $N^3$ for SP-BICs, which is of vital importance for realizing quasi-BICs in a compact PhC. Microwave experiments are performed to validate the theoretical results. Our results provide a paradigm shift for manipulating the physical properties quasi-BICs in finite PhC structures, which would facilitate various applications, including but not limited to low threshold lasing, wireless power transfer and high figure of merit sensing etc.
\end{abstract}


\maketitle

\textit{Introduction.---} Different from whispering gallery mode \cite{chen2017exceptional,jiang2017chaos,microcavity photonics} and photonic crystal defect mode \cite{nakadai2022electrically}, BIC enables the counter-intuitive localization of wave in a continuum spectrum whose wave functions are generally extended \cite{hsu2016bound}. It is a very generalized physical phenomenon that has been observed in various wave systems, such as optics, acoustics and mechanics \cite{hsu2016bound, sadreev2021interference,chen2016mechanical,Azzam2020,koshelev2019meta}. 
Leveraging by its theoretical infinite $Q$ factor, many interesting physical mechanisms and promising applications have been explored and demonstrated, such as conservation of topological charges for polarization \cite{zhen2014topological}, manipulation of BICs in momentum space
\cite{jin2019topologically,wang2020generating,yoda2020generation,Bulgakov PRL}, lasing \cite{huang2020ultrafast,song2020coexistence,hwang2021ultralow,ssss}, chiroptical response \cite{chen2022can,zhang2022chiral,shi2022planar}, biosensing \cite{yesilkoy2019ultrasensitive} and enhancement of Smith-Purcell radiation \cite{yang2018maximal}, just to name a few. Specifically, PhC resembles a promising platform for tailoring the properties of BIC, though genuine BIC only exists in an infinite PhC \cite{hsu2016bound}. However, any BICs inevitably turns into quasi-BICs with finite $Q$ factors in practice. Therefore, there is a general quest to boost the $Q$ factors of quasi-BICs in compact PhC structures considering the promising prospects of harnessing the functionalities of BICs. To date, one straightforward method for boosting the $Q$ factor of quasi-BIC relies on increasing numbers of units for the finite PhC\cite{liu2019high, Sadrieva2019, bulgakov2019high}, which is fundamentally limited by material losses \cite{Sadrieva2019} and structural fluctuations during fabrication \cite{Ni2017}. Tailoring the morphology of unit cells for a finite PhC is also proved to be quite effective in achieving high-$Q$ quasi-BICs of PhC structures with a small footprint \cite{koshelev2018asymmetric, liu2019high}. In particular, merging multiple BICs in momentum space is another effective and sophisticated mean to boost the $Q$ factors of quasi-BICs \cite{jin2019topologically, kang2021merging,chen2022observation}, which is independent of the mechanism of achieving quasi-BICs by avoided crossing of resonances (ACR) \cite{rybin2017,bogdanov2019bound}.

\begin{figure}[htb!]
	\begin{center}
		\includegraphics[width=1\linewidth]{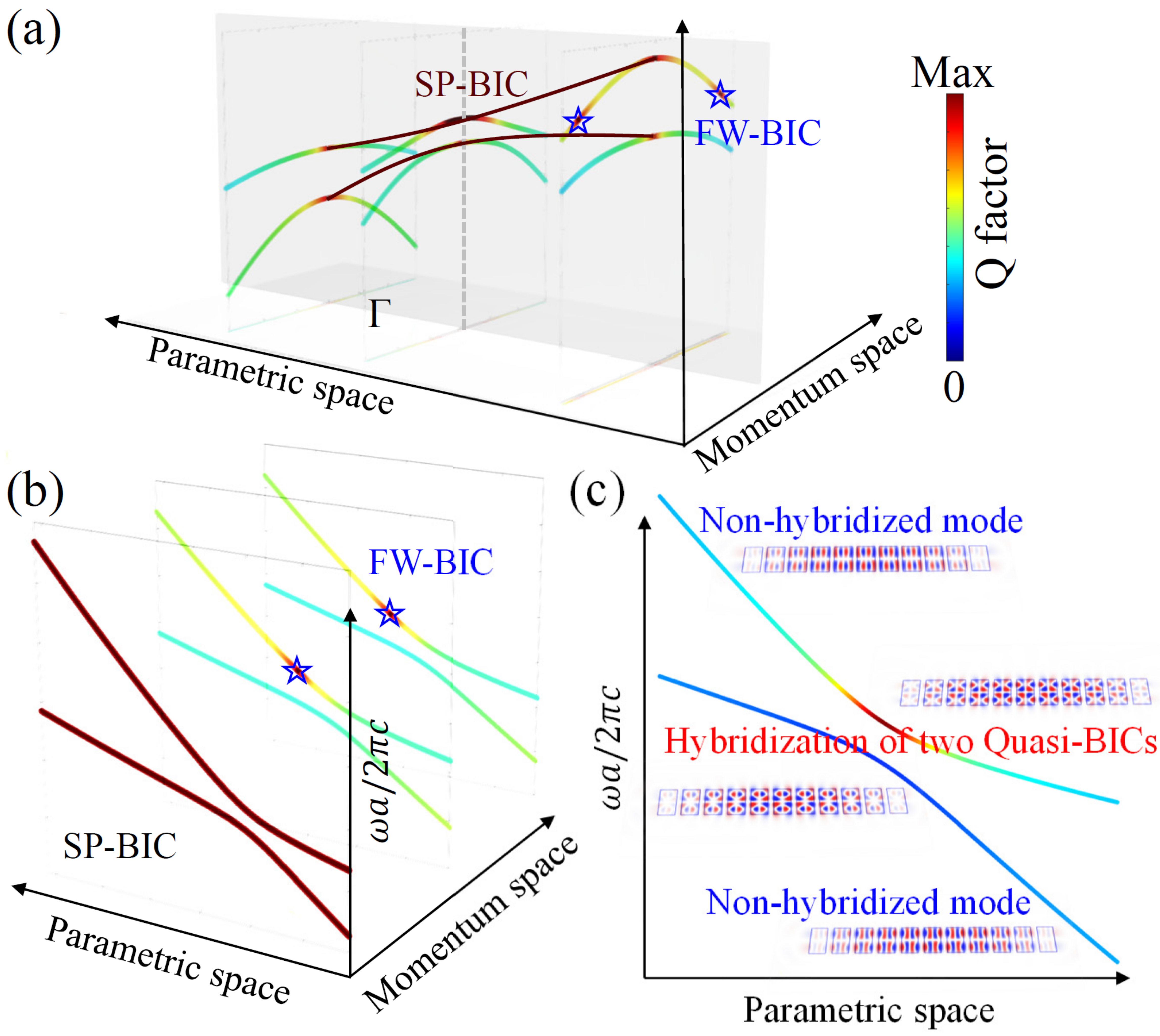}
		\caption{(a) Schematic for the hybridization of two SP-BICs of an infinite PhC in parametric space together with merging BICs in momentum space. The $Q$ factors of eigenmodes are indicated by color code. The FW-BICs at off-$\Gamma$ points are outlined by the stars. (b) The off-$\Gamma$ BICs are because of the interband coupling in parametric space. (c) Schematic for the hybridization of two quasi-BICs in a finite PhC, where it occurs near the avoided crossing region. Typical distributions of electric field ($Re(E_{y})$) are shown in the insets.}
		\label{fig:fig1}
	\end{center}
\end{figure}

However, all these methods break down when the non-radiative loss of the finite PhC surpasses the radiative loss. As a result, the non-radiative loss will impose an upper limit of $Q$ factor for a PhC with a given number of units, when the radiative loss of quasi-BIC is smaller than the nonradiative one \cite{Sadrieva2019}. This fact essentially pinpoints the importance of asymptotic behavior of quasi-BICs' $Q$ factors over the number of periodic unit $N$, i.e. $Q(N)\sim$ $N^{\eta}$. Therefore,
exploring new ability to boost the $Q$ factor of a compact PhC by enlarging the asymptotic factor $\eta$ becomes very important, where the radiative $Q_{r}$ should surpass the upper bound set by its non-radiative $Q_{nr}$. It is also appealing to develop an elegant mechanism which is also fully compatible with the mechanism of merging BICs.

\begin{figure*}[htb!]
	\begin{center}
		\includegraphics[width=\linewidth]{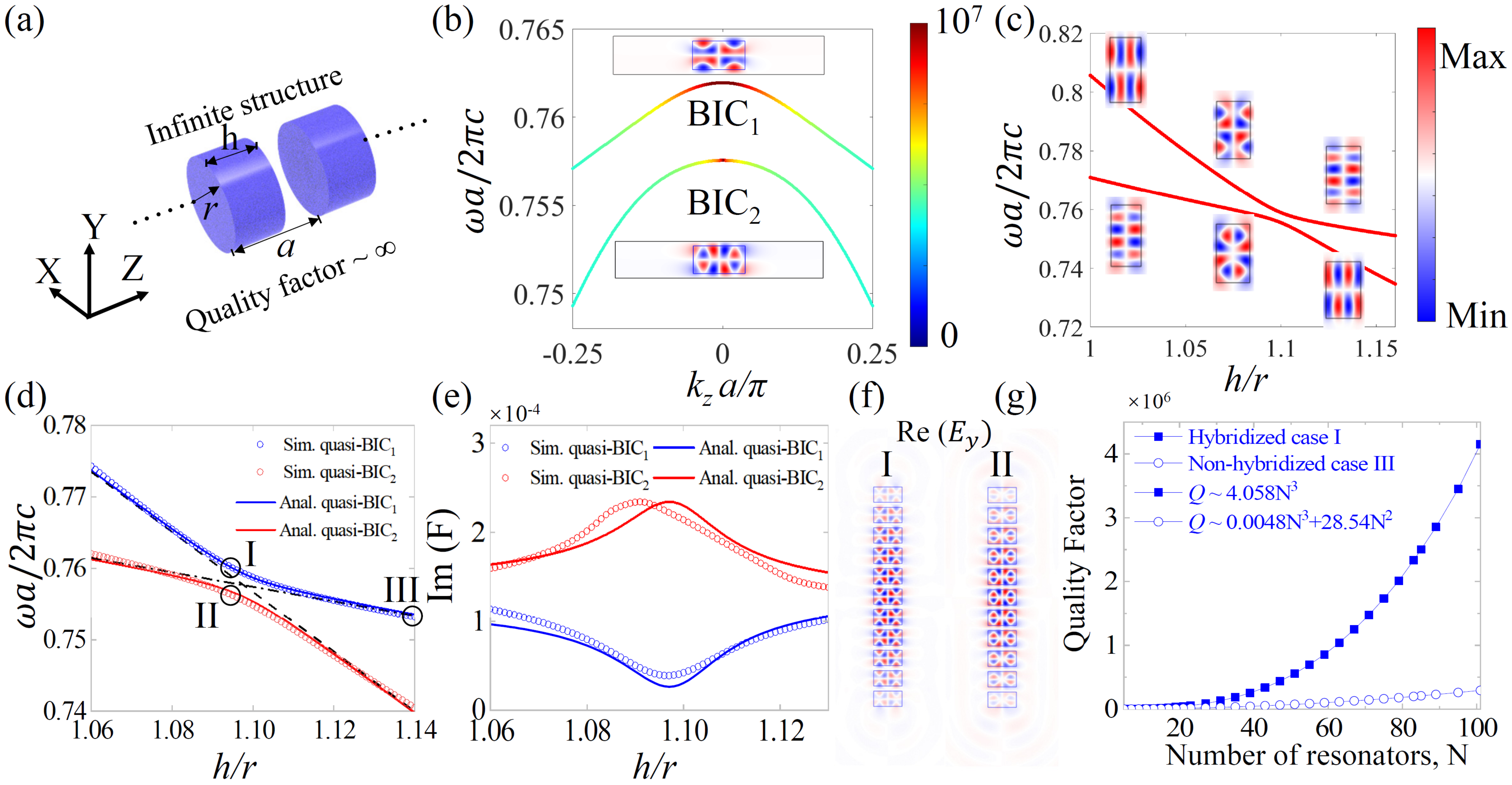}
		\caption{(a) Schematic of a 1D PhC. The PhC is formed by coupled dielectric cylindrical resonators with a radius $r$ and a height $h$. The relative permittivity $\varepsilon_{r}$ is 9.6 and the period $a$ is $r$/0.688. (b) The modal dispersion and corresponding $Q$ factors (color coded) of the structure ($h$/$r$ = 1.0908) when merging BICs occurs in momentum space. (c) The eigenfrequencies of two SP-BICs for infinite PhCs with different $h$/$r$. The corresponding distributions of electric field (Re($E_{y}$)) on the X-Z plane are shown in the inset. (d) and (e) The real and imaginary parts of eigenfrequencies of two quasi-BICs for the finite PhC consisted of 11 units obtained by simulations and theoretical model. (f) The corresponding distributions of electric field (Re($E_{y}$)) for modes I and II. (g) The $Q$ factors of hybridized ($h$/$r$ = 1.0908, upper branch) and non-hybridized ($h$/$r$ = 1.14, upper branch) modes as a function of resonator number (N) for the finite PhCs. The fitting formulas are also provided.}
		\label{fig:fig2}
	\end{center}
\end{figure*}

In this letter, we theoretically propose and experimentally demonstrate a new concept of super quasi-BIC, as schematically shown in Fig. \ref{fig:fig1}. This concept is mediated by the avoided crossing of BICs in parametric space [see the gray frame in Fig. \ref{fig:fig1} (a)] together with the merging BICs in momentum space, where the off-$\Gamma$ BICs are Friedrich-Wintgen BIC (FW-BIC) originated from interband coupling [Fig. \ref{fig:fig1} (b)]. More importantly, the avoided crossing of two quasi-BICs is capable of enhancing the $Q$ factor for one of them, as schematically shown in Fig. \ref{fig:fig1} (c). Microwave experiments are performed to consolidate the theoretical results.

\textit{Avoided crossing of BICs and quasi-BICs.---}
Without loss of generality, we consider the simplest one-dimensional (1D) PhC shown in Fig. \ref{fig:fig2} (a), where BICs only exist if the structures are periodic
\cite{vincent1979corrugated,kim2019optical,shipman2005resonant,hsu2013observation,bulgakov2014bloch,
bulgakov2017bound,sadrieva2017transition,Gao2019,lee2019band,Maslova2021,blaustein2007guiding,polishchuk2017guided,bulgakov2017light,asenjo2017exponential}.
The 1D PhC consists of infinite coaxial coupled dielectric resonators, which is structurally similar to fiber Bragg gratings \cite{Gao2019}. The unit cell of the PhC is composed of a cylindrical resonator with a radius of $r$, a height of $h$, a relative permittivity of $\varepsilon_{r}$ = 9.6 and a period of $a$ = $r$/0.688. The background material is air. This PhC supports BICs localized at the X-Y plane \cite{bulgakov2017bound, kim2019optical}, which are classified according to the irreducible representations of rotations around the symmetry axis $C_{\infty}$ specified by a
azimuthal index $m$. The case of $m$ = 0 for the SP-BICs is considered \cite{bulgakov2017bound}.
The modal dispersion and the corresponding $Q$ factors are calculated by the finite-element method (FEM) (Supplemental Material \cite{SI}, Sec. I).
The corresponding results are shown in Fig. \ref{fig:fig2} (b), where there are two BICs in the first diffraction continuum. Based on the electric field distributions ($Re(E_{y})$) shown in the insets, it can be concluded that they are SP-BICs originated from symmetry incompatibility with the continuum. By changing the structural parameters $h/r$, two SP-BICs will undergo avoided crossing that brings a new degree of freedom to tailor the SP-BICs. As can be seen from Fig. \ref{fig:fig2} (c), the normalized frequencies of two SP-BICs are close but avoided crossing from each other near $h$/$r$ = 1.09, which is because of coupling via the next diffraction continua. The corresponding mode profiles at six typical positions of two branches are also shown in the insets. It can be seen that the SP-BICs are strongly hybridized near the avoided crossing region. Because this interaction takes place at the $\Gamma$ point, all hybridized modes remain SP-BICs \cite{chen2022can}. At the same time, merging BICs is also achieved for the BIC$_{1}$ (Supplemental Material \cite{SI}, Sec. II), enabling the unprecedented tailoring of BICs in both parametric and momentum spaces.

\textcolor{black}{In order to qualitatively describe the concurrence of avoided crossing of BICs and merging BICs, the following effective non-Hermitian Hamiltonian can be introduced \cite{sadreev2021interference} (see details in Supplemental Material \cite{SI}, Sec. III). As can be seen from Fig. \ref{fig:fig2} (b), the dispersion over $k_{z}$ shows topical parabolic dependence. As a result, the real parts of eigenfrequencies corresponding for two bands ($\Omega_{1,2}$) can be approximately described by Eq. (\ref{om12}), where $\omega_{1,2}$ specify the real parts of eigenfrequencies for two SP-BICs with different aspect ratios $h$/$r$ and $k_{z}$ represents the wave vector, respectively.
\begin{equation}\label{om12}
\Omega_{1,2}(k_z)=\omega_{1,2}(h/r)+a_{1,2}k_z^2
\end{equation}}
Therefore, the general two-level description of effective non-Hermitian Hamiltonian is 
\begin{equation}\label{Hefff}
\operatorname{H_{eff}}=\left(\begin{array}{cc}
\varepsilon+ak_{z}^2 & u \\
u & -\varepsilon-ak_{z}^2
\end{array}\right)-i\left(\begin{array}{cc}
G_1 & \sqrt{G_1 G_2} \\
\sqrt{G_1 G_2} & G_2
\end{array}\right)
\end{equation}
\begin{equation}\label{G12}
G_{1,2}(k_z)=\gamma_{1,2}+b_{1,2}k_z^2
\end{equation}
Here $ \epsilon=\omega_1(h/r)-\omega_2(h/r)$ is the distance between two bands at $k_z$ = 0 while the parameter $a=a_1-a_2$ describes the distance at $k_z\neq 0$. These distances depend on the geometry parameter of the PhC, or on the aspect ratio in our case.
$u$ and $\sqrt{G_1 G_2}$ are the near-field and radiation coupling, respectively. $G_{1,2}$ are the decay rates of the modes. $\gamma_{1,2}$ represent the decay rates at $k_z$ = 0 and $b_{1,2}$ are the coefficients of the parabolic dependence. When the following condition is satisfied, 
\begin{equation}\label{con}
u (G_1-G_2)=2\sqrt{G_1 G_2}(\varepsilon+ak_{z}^2)
\end{equation}the FW-BIC will appear. Based on Eq. (\ref{con}), avoided crossing of BICs and merging BICs will occur when $k_z$ = 0 and $\gamma_{1,2}$ = 0 (Supplemental Material \cite{SI}, Sec. III).

This mechanism also provides a promising way to manipulate the quasi-BIC in a compact PhC, whose $Q$ factor is determined by the longitudinal leakage along $Z$ axis ($Q_{\parallel}$) and the transverse leakage ($Q_{\perp}$), respectively. 
\begin{equation}\label{QTag}
    \frac{1}{Q}=\frac{1}{Q_{\perp}}+\frac{1}{Q_{\parallel}}.
\end{equation}
The longitudinal leakage is because of the loss from the ends of 1D PhC and it has an universal dependence of $Q_{\parallel} \sim N^3$ \cite{Sadrieva2019,Maslova2021,blaustein2007guiding,polishchuk2017guided,bulgakov2017light,asenjo2017exponential,bulgakov2019high}. By assuming that the transverse radiation leakage per unit length of the PhC is similar to the infinite case and considering a quasi-BIC supported by $N$ resonators with the minimal Bloch wave vector $k_z$ = $\pi/Na$, the power that radiates perpendicular to $Z$ axis can be evaluated (Supplemental Material \cite{SI}, Sec. IV). Then, the asymptotic behavior for the $Q_{\perp}$ factor govern by the transverse leakage can be obtained according to the definition of $Q$ factor:
\begin{equation}\label{QTag1}
\frac{1}{Q_{\perp}} \approx \frac{1}{Q_{0}}\left\{\frac{\alpha}{N}+\frac{\beta}{N^{3}}\right\}^{2}
\end{equation}
As can be seen from Eqs. (\ref{QTag}) and (\ref{QTag1}), the asymptotic behavior of $Q$ factor becomes $Q \sim N^2$ for the general case of SP-BIC (i.e. $\alpha, \beta \neq$ 0), where the transverse leakage is dominated. This case is similar to the reported scenario of SP-BICs without hybridization \cite{Sadrieva2019,bulgakov2019high} (Supplemental Material \cite{SI}, Sec. V). However, the coefficient $\alpha$ could become zero by tuning the geometry parameters because of destructive interference (Supplemental Material \cite{SI}, Sec. IV), resulting in the asymptotic behavior of $Q \sim N^3$. There is also an intermediate case that $\alpha$ is close but not exactly zero, where both $N^2$ and $N^3$ terms are contributed to the asymptotic behavior. Merging BICs without ACR could be one typical example (Supplemental Material \cite{SI}, Sec. VI). Therefore, ACR becomes a new degree of freedom to exclusively preserve the $N^3$ asymptotic behavior.

Considering a finite 1D PhC with 11 coupled resonators, the SP-BIC$_{1}$ and BIC$_{2}$ shown in Fig. \ref{fig:fig2} (b) change to quasi-BICs with finite $Q$ factors. Especially, two quasi-BICs can also hybridize with each other by changing the aspect ratio $h/r$ of the resonator, leading to the ACR intuitively similar to the infinite case for the real parts of eigenfrequencies, as shown in Fig. \ref{fig:fig2} (d). However, there is essential difference for the imaginary parts of eigenfrequencies, where they can be effectively manipulated near the ACR, as shown in Fig. \ref{fig:fig2} (e). As a result, the $Q$ factors of modes are substantially modified during the hybridization, where the imaginary part of eigenfrequency for mode I in the upper branch is effectively reduced. This property pinpoints the importance of studying the asymptotic property in finite PhC. The corresponding distributions of electric field ($Re(E_{y})$) for modes I and II are shown in Fig. \ref{fig:fig2} (f). Compared with the infinite case shown in Fig. \ref{fig:fig2} (c), the distributions of electric field within the unit cell of both PhCs are qualitatively similar, indicating the quasi-BICs under consideration are originated from SP-BICs shown in Fig. \ref{fig:fig2} (c). 

According to the fitting results, the asymptotic behavior of the non-hybridized case (mode III) far away from the ACR is dominantly proportional to $N^2$, as shown in Fig. \ref{fig:fig2} (g). While the $Q$ factor of the hybridized case increases much more rapidly compared with the non-hybridized case. More importantly, the asymptotic property of the hybridized quasi-BIC (mode I) at the upper branch exclusively becomes $N^3$, corresponding to the case of $\alpha$ = 0 as indicated in Fig. \ref{fig:fig2} (g) (Supplemental Material \cite{SI}, Sec. V). 
Such a transition is very important for boosting the $Q$ factors of quasi-BICs which becomes saturated together with the increasing of $N$ for lossy materials (Supplemental Material \cite{SI}, Sec. V). 
These results suggest that the asymptotic behavior of the quasi-BIC's $Q$ factor over the number of resonators can be effectively manipulated by using the concurrence of avoided crossing of quasi-BICs and merging BICs. As a result, the super quasi-BIC can be realized in a compact PhC.

By generalizing the effective non-Hermitian Hamiltonian of Eq. (\ref{Hefff}) to include the finite size effect, reasonable agreement can be achieved between the results of eigensolutions and the model, as shown in Fig. \ref{fig:fig2} (d) and (e). There is still discrepancy between the two-level model and the eigensolutions. This is because there is another resonance of the finite PhC nearby, which is beyond the two-level model (Supplemental Material \cite{SI}, Sec. III).

\begin{figure}[htb!]
	\begin{center}
		\includegraphics[width=\linewidth]{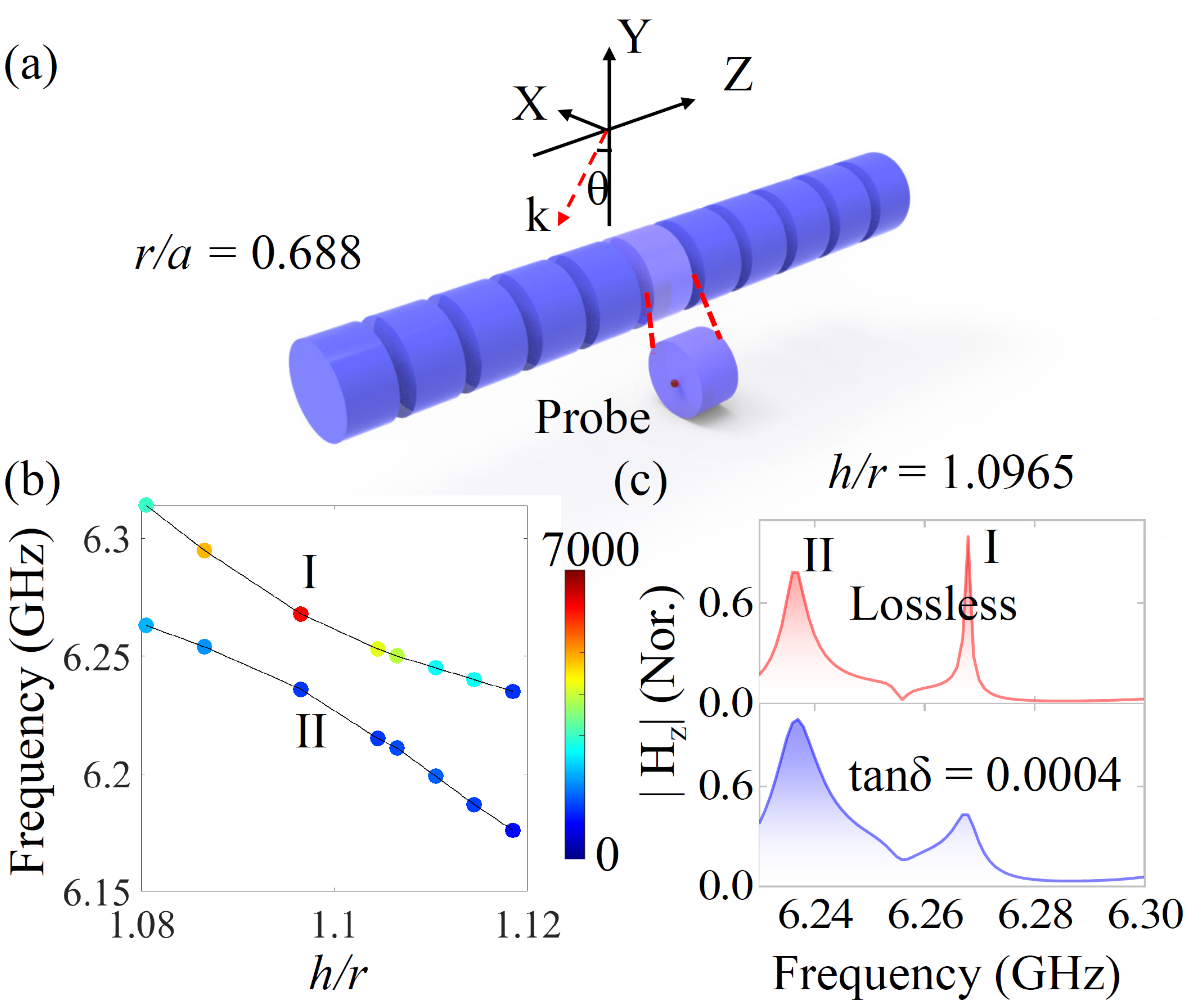}
		\caption{(a) The quasi-BICs are excited by an inclined plane wave with an $E_{x}$ polarization, where its k-vector has an angle $\theta$ with respect to Y axis. (b) The frequencies and the corresponding $Q$ factors (color coded) of quasi-BICs for different $h$/$r$. The magnetic field components are probed at the position marked by a red sphere in (a). (c) Normalized ($\vert H_{z}\vert$) as a function of excitation frequencies at the case of $h$/$r$ = 1.0965 for lossy (tan $\delta$ = 0.0004) and lossless cases, respectively.}
		\label{fig:fig3}
	\end{center}
\end{figure}

We further calculate the plane wave scattering of a PhC consisted of 11 units by using the FEM (Supplemental Material \cite{SI}, Sec. I), as shown in Fig. \ref{fig:fig3} (a). The quasi-BICs modes can be excited by an inclined plane wave ($\theta$ = 5$^o$) with $E_{x}$ polarization. 
The spectrum responses of magnetic field component $\vert H_{z}\vert$ at the position marked in Fig. \ref{fig:fig3}
(a) are used to extract the resonant frequencies and their corresponding $Q$ factors for different aspect ratios $h$/$r$, where the results are shown in Fig. \ref{fig:fig3} (b). As can be seen from these results, there is a typical ACR in the considered frequency range. The highest $Q$ factor ($\sim$ 7000) appears at $h$/$r$ = 1.0965, as shown in Fig.\ref{fig:fig3} (c). Both lossless and lossy (tan $\delta$ = 0.0004) cases are considered. These results are similar to the eigensolutions presented in Fig.\ref{fig:fig2} (d) and (e).
It is the avoided crossing of two quasi-BICs that results in the enhancement of $Q$ factor for the quasi-BIC I. The mechanism for boosting the $Q$ factor by the ACR can also be elucidated by using electromagnetic multipole decomposition \cite{grahn2012electromagnetic}. The enhancement of $Q$ factor for the quasi-BIC is related to but different from electromagnetic multipolar conversions from lower to higher orders found in the avoided crossing of Mie resonances \cite{chen2019multipolar} (Supplemental Material \cite{SI}, Sec. VII), where merging BICs cannot be achieved.

\begin{figure}[htb!]
\begin{center}
\includegraphics[width=\linewidth]{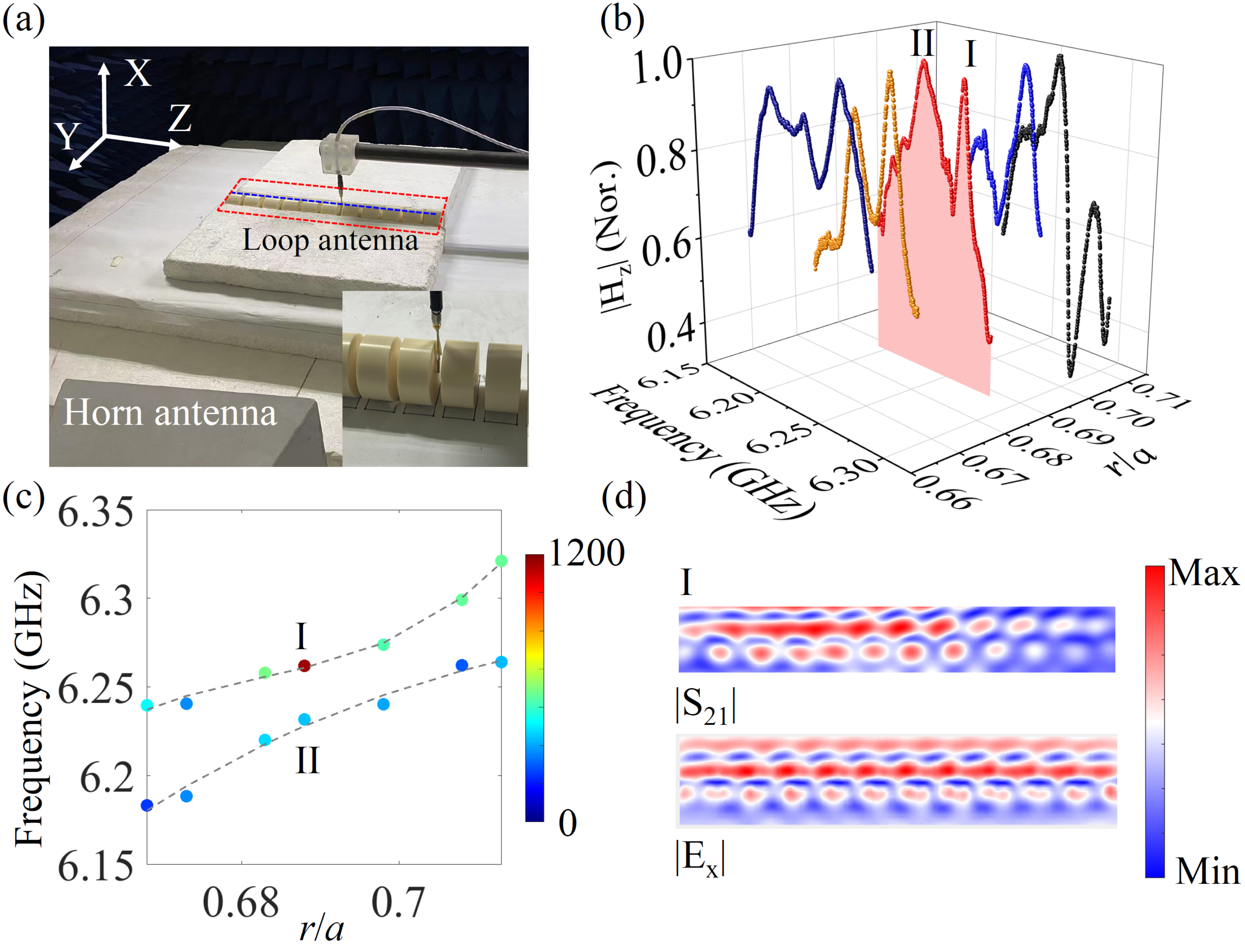}
\caption{(a) The experimental setup. The position of loop antenna is shown by the inset. (b) The experimental acquired $\vert S_{21}\vert$ measured by a loop antenna under the excitation of the horn antenna for PhCs with different $r$/$a$. (c) The corresponding extracted frequencies and $Q$ factors (color coded) of quasi-BICs. The fitting curves are also provided. (d) The distribution of $\vert S_{21}\vert$ measured by a monopole antenna at the red dashed box outlined in (a) and the corresponding simulation results $\vert E_{x}\vert$ above the PhC.}
\label{fig:fig4}
\end{center}
\end{figure}

\textit{Experimental demonstrations.---} Microwave experiments are performed to validate the theoretical results. In experiment, it is inconvenient to vary the aspect ratio $h$/$r$ of the PhC to control the hybridization of two quasi-BICs. Instead, it is achieved by changing the ratio between radius and lattice constant $r$/$a$, where similar results can be achieved. A PhC composed of 11 coupled dielectric resonators fabricated by alumina ceramics ($\varepsilon_{r}$ = 9.6, $\tan\delta\approx4\times 10^{-4}$, $r$ = 25 mm, h = 1.096 $r$) is considered, which supports a super quasi-BIC mentioned in Fig. \ref{fig:fig1} (Supplemental Material \cite{SI}, Sec. VIII). 

The experimental setup for performing near-field measurement of electromagnetic field is shown in Fig. \ref{fig:fig4} (a). Normalized $\vert S_{21} \vert$ of an antenna as functions of frequency and positions can be measured. The magnetic field components $\vert H_{z}\vert$ probed within the coupled resonators for different $r$/$a$ are shown in Fig. \ref{fig:fig4} (b) (see Supplemental Material \cite{SI}, Sec. I for experimental details). As can be seen qualitatively from this figure, the $Q$ factors of modes associated with the quasi-BIC $I$ are substantially enhanced around $r$/$a$ = 0.688. Quantitatively, the frequencies and the corresponding $Q$ factors (color coded) of these resonances are presented in Fig. \ref{fig:fig4} (c). It can be found that the avoided crossing appears exactly near $r$/$a$ = 0.688 where the resonance $I$ possesses the highest $Q$ factor. 
The two-dimensional distribution of $\vert S_{21} \vert$ at the quasi-BIC $I$ measured at the red dashed box marked in Fig. 4 (a) are shown in Fig. 4 (d), where the corresponding simulation result is also provided. Similar results between the calculated and measured ones can be achieved, which confirms that the enhancement of $Q$ factors in this finite PhC relies on the hybridization of two quasi-BICs in parametric space and the merging BICs in momentum space. The discrepancy between the simulation and experimental results might be caused by the non-flat wavefront produced by the horn antenna (Supplemental Material \cite{SI}, Sec. IX). It should be noted that the $Q$ factor achieved in the experiment can be further enhanced by applying material with lower loss \cite{liu2019high} and modifying the boundary \cite{chen2022observation}. 

\textit{Conclusions and outlooks.---}
In summary, we theoretically study and experimentally demonstrate a new concept of super quasi-BIC, where the avoided crossing of quasi-BICs in parametric space and merging BICs in momentum space are achieved simultaneously. Therefore, the asymptotic behavior of $Q$ factor over the unit number $N$ of PhC is exclusively determined by $N^3$, which is very important for boosting the $Q$ factor of SP quasi-BICs in a compact and lossy PhC. Microwave experiments further verify the theory results. The unprecedented manipulation of quasi-BICs in both parametric and momentum spaces provides a paradigm shift for manipulating quasi-BICs and opens new possibility for localizing electromagnetic wave within the radiative continuum. It is anticipated that our findings could pave the way for achieving a higher $Q$ factor quasi-BIC in a more compact PhC, facilitating the miniaturization of the electromagnetic resonators and substantially enriching the physics of BICs.

\begin{acknowledgments}
Zhanyuan Zhang and Yi Xu would like to thank the suggestions from Dr. Meng Kang and Prof. Chengbo Mou. This work was supported by National Natural Science Foundation of China (62222505), Russian Science Foundation with Grant number 22-12-00070 and the Guangdong Introducing Innovative, Entrepreneurial Teams of "The Pearl River Talent Recruitment Program" (2019ZT08X340).
\end{acknowledgments}


\end{document}